\documentclass{sf2a-conf2014}
\usepackage{graphicx}
\usepackage{hyperref}
\usepackage[]{natbib}  
\usepackage{epstopdf}
\usepackage{color}

\def\BibTeX{{\rm B\kern-.05em{\sc i\kern-.025em b}\kern-.08em
    T\kern-.1667em\lower.7ex\hbox{E}\kern-.125emX}}
\bibpunct{(}{)}{;}{a}{}{,}  

\begin{document}

\TitreGlobal{SF2A 2014}


\title{Probing the Galactic center with X-ray polarimetry}

\runningtitle{Polarimetric prospects on the Galactic center}

\author{F.~Marin}\address{Astronomical Institute of the Academy of Sciences, Bo{\v c}n\'{\i} II 1401, 
			   CZ-14100 Prague, Czech Republic}
			  
\author{V. Karas$^1$}

\author{D. Kunneriath$^1$}

\author{F. Muleri}\address{INAF/IAPS, Via del Fosso del Cavaliere 100, I-00133 Roma, Italy}

\author{P. Soffitta$^2$}				

\setcounter{page}{237}


\maketitle


\begin{abstract}
The Galactic center (GC) holds the closest-to-Earth supermassive black hole (SMBH), which makes it the best laboratory to study the close environment 
of extremely massive compact objects. Polarimetry is sensitive to geometry of the source, which makes it a particularly suitable technique to probe 
the medium surrounding the GC SMBH. The detection of hard X-ray spectra and prominent iron K$\alpha$ fluorescence features coincident with localized
gas clouds (e.g. Sgr~B2, Sgr~C) is known for nearly twenty years now and is commonly associated with a past outburst of the SMBH whose radiation is 
reprocessing onto the so-called ``reflection nebulae''. Since scattering leads to polarization, the re-emitted signal from the giant molecular clouds 
in the first 100~pc of the GC is expected to be polarized. X-ray polarization measurement is thus particularly adapted to probe the origin of the 
diffuse X-ray emission from the GC reflection nebulae and reveal the past activity of the central SMBH. In this research note, we summarize the 
results from past and current polarimetric simulations in order to show how a future X-ray polarimeter equipped with imaging detectors could improve 
our understanding of high-energy astrophysics.
\end{abstract}

\begin{keywords}
Galaxy: center, Galaxy: nucleus, Polarization, Radiative transfer, Scattering
\end{keywords}


\section{Introduction}
Using the ART-P telescope on-board of \textit{Granat}, \citet{Sunyaev1993} were the first to image the Galactic center (GC) from the soft (2.5~keV)
to the hard (22~keV) X-ray band. The discovery of a source characterized by a spherical morphology in the soft band, and by an extended (i.e. elongated 
along the Galactic plane) shape in the hard X-ray range, lead \citet{Sunyaev1993} to suggest that part of the hard X-ray emission could be due to 
Compton-scattered photons originating from a nearby compact source and reprocessed by dense molecular clouds. Later, X-ray images and spectra of the GC
obtained by \citet{Koyama1996}, using the Advanced Satellite for Cosmology and Astrophysics (\textit{ASCA}), revealed fluorescent K$\alpha$ emission 
lines from cold iron atoms in Sgr~B2. Associated with a hard X-ray spectral slope, \citet{Koyama1996} estimated that the diffuse thermal emission
could be due to a past intense irradiation from Sgr~A$^*$, the central supermassive black hole (SMBH) that lies in the GC. Such an argument is in 
agreement with the predictions of \citet{Sunyaev1993} and further detections from other extended gas cloudlets tend to corroborate this idea. 
In particular, \citet{Murakami2001} presented the first \textit{ASCA} evidence of diffuse hard X-ray emission associated with strong 6.4~keV fluorescence 
line and large absorption from the Sgr~C complex. Hard X-ray continuum and Fe~K$\alpha$ emission have also been found in several other GC molecular 
clouds: Sgr~B1 \citep{Koyama2007}, M0.74-0.09 \citep{Koyama2007}, G0.11-0.11 \citep{Ponti2010}, M0.74-0.09 \citep{Nobukawa2011}, M1 and M2 
\citep{Ponti2010}, the Arches cluster \citep{Yusef-Zadeh2002} and the molecular complex called the Bridge \citep{Bamba2002}. 

To uncover the nature of the hot diffuse X-ray emission and test the hypothesis of the flaring theory, \citet{Churazov2002} postulated that an X-ray 
mission equipped with a state-of-the-art polarimeter would be necessary. Indeed, if past radiation from Sgr~A$^*$ is reprocessed by extended, distant reflection 
nebulae, the resulting X-ray emission should be polarized. In order to lay the groundwork for a future polarimetric explorer, several studies have been done
\citep{Churazov2002,Matt2010,Marin2014}, estimating the polarization degree and the orientation of the polarization position angle that a potential mission could detect. 
Hence, in this research note, we summarize the work accomplished so far to evaluate the net X-ray polarization emerging from the largest and brightest 
(i.e. easier to detect) GC molecular clouds.

\section{Evaluating the polarization emerging from the GC}
So far, there have been two main publications exploring the resulting X-ray polarization emerging from the GC. The first one, from \citet{Churazov2002},
focused on the reprocessing molecular cloud Sgr~B2 and estimated the net soft X-ray polarization signature of this cloud. The second paper, by 
\citet{Marin2014}, modeled the 2$^\circ~\times~$2$^\circ$ ($\sim$~288$~\times~$288~pc; at 8.5~kpc 1$^{\prime\prime}$ $\approx$ 0.04~pc) central 
region of the Milky Way at energies higher than 8~keV and included polarization maps. We now review the main results from the two complementary 
publications.

\subsection{Sgr~B2 in the soft X-ray band}

\citet{Churazov2002} investigated the polarization emerging from a single cloud of radius 10~pc representative of the reflection nebulae Sgr~B2.
In their simulation, the spherical gaseous medium was filled with neutral, solar abundance matter and had a Thomson optical depth of 0.5. It was
irradiated by a steady beam of unpolarized photons with energy ranging from 2 to 8~keV. Photoelectric absorption, fluorescent emission and Compton
scattering by bound electrons were included and the molecular cloud was located at 100~pc from the continuum source, which is representative of the 
known projected distance from Sgr~A$^*$ \citep{Murakami2001}. However, since the three-dimensional parametrization of the GC is unknown, 
the reflecting nebula can be shifted from the Galactic plane but still conserve the same radial projections. To investigate this effect,
\citet{Churazov2002} located their cloud model at two different positions: at the same distance and 100~pc away from the observer than the 
emitting region.

\begin{figure*}[!t]
\centering
   \includegraphics[trim = 0mm 2mm 0mm 4mm, clip, width=16cm]{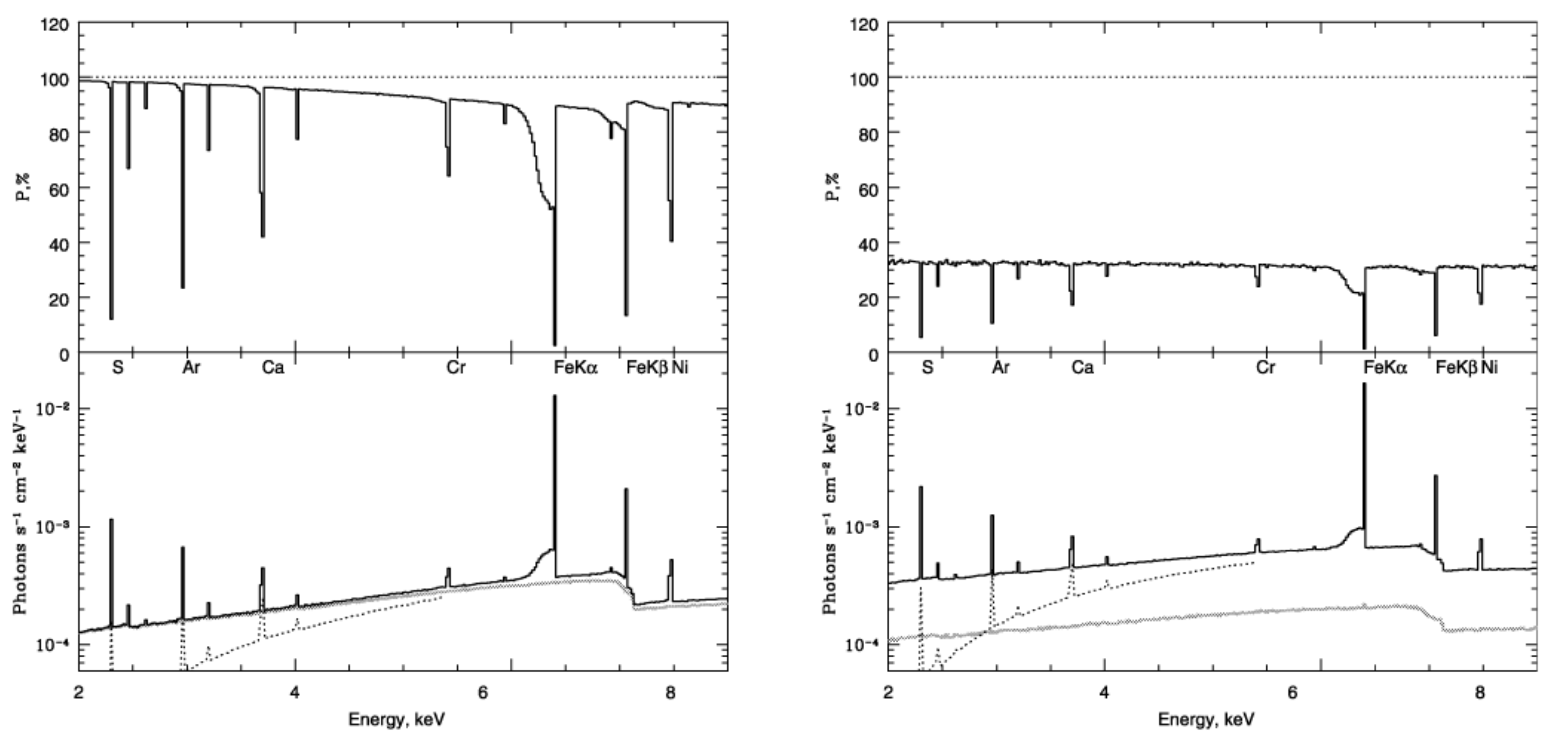}
   \caption{Reflected energy spectrum and polarization degree of a spherical gas cloud 
	    illuminated by a distant continuum source. \textit{Top}: Polarization degree $P$, 
	    \textit{bottom}: intensity spectra. The left column is representative of a gas 
	    clump situated at the same distance from the observer as the source,
	    while the right column presents results for a cloud situated 100~pc away. 
	    Spectra are taken from \citet{Churazov2002}.}
  \label{Fig:Churazov}
\end{figure*}

Their work is summarized in Fig.~\ref{Fig:Churazov} and the general result is that the reflected radiation should be highly polarized ($>$~30\%) 
with a direction of polarization normal to the scattering plane. The amount of $P$ is a function of the three-dimensional position of the cloud,
decreasing when Sgr~B2 departs from the Galactic plane. Local dilution of the polarization degree by fluorescent photons (being unpolarized)
is symptomatic of the composition of the cloud, and the slow but steady decrease of $P$ with energy is due to multiple scatterings. 

~\

Thus, using a simple model, \citet{Churazov2002} proved that any detection of polarized X-ray emission from the reflection nebulae around Sgr~A$^*$ 
would give constraints on the morphology, composition and position of the scattering clouds.

\subsection{Broadband GC polarization imaging}

The results obtained by \citet{Churazov2002} are valid in the 2 -- 8~keV band for small hydrogen column densities (n$_{\rm H}~<$~10$^{22~}$cm$^{-2}$). 
At larger n$_{\rm H}$, the emission below 5~keV is expected to be highly suppressed. In addition to that, past X-ray observations 
\citep{Koyama1986,Koyama1989} revealed the presence of a diffuse plasma emission toward the GC that may additionally dilute the polarization signal 
below 7~keV. Hence, to avoid most of the dilution by the GC plasma emission and extend the soft X-ray simulations achieved by \citet{Churazov2002},
8 -- 35~keV modeling has been undertaken by \citet{Marin2014}.

\begin{figure*}[!t]
\centering
   \includegraphics[trim = 0mm 3mm 0mm 5mm, clip, width=10cm]{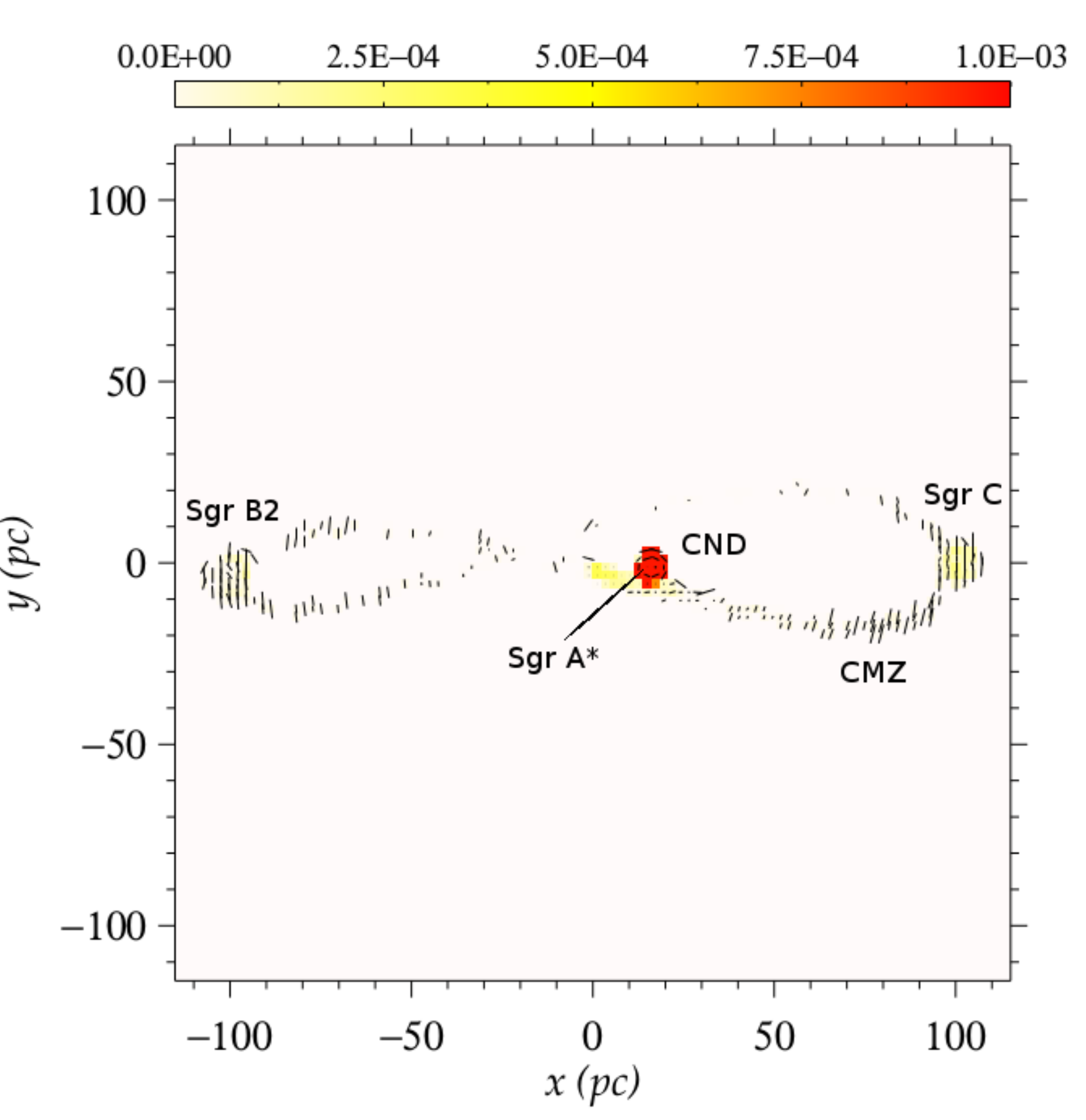}
   \caption{Integrated 8 -- 35~keV model image of the polarized flux, $PF/F_{\rm *}$,
	    for the 2$^\circ~\times~$2$^\circ$ region around the GC. $PF/F_{\rm *}$ is 
	    color-coded, with the color scale shown on top of the image (in arbitrary units).
	    $P$ and $\psi$ are represented by black bars drawn in the center 
	    of each spatial bin. A vertical bar indicates a polarization angle of 
	    $\psi$~=~90$^\circ$ and a horizontal bar stands for an angle
	    of $\psi$~=~0$^\circ$. The length of the bar is proportional to 
	    $P$. The polarization map is taken from \citet{Marin2014}.}
  \label{Fig:Marin}
\end{figure*}

In our recent work, \citet{Marin2014} took into account the dense and warm environment around the central few parsecs around Sgr~A$^*$ (known as the 
circumnuclear disc, CND); a cold dusty structure in the shape of a continuous chain of irregular clumps representative of the central 
molecular zone (CMZ, \citealt{Molinari2011}); the two bright, complex reflection nebulae Sgr~B2 and Sgr~C (located at the two extrema of the 
CMZ); and a continuum source displaced by $\sim$~22~pc in projection from the center of the model toward the Western galactic longitude 
(to be consistent with the shifted gas distribution within the CMZ \citep{Molinari2011}. 

The polarization map presented in Fig.~\ref{Fig:Marin} is extracted from \citet{Marin2014}. It is found that the polarized flux, $PF/F_{\rm *}$, traces 
the overall shape of the GC, emphasizing the $\infty$-shaped CMZ. $PF/F_{\rm *}$ reaches a maximum at the location of the non-axisymmetric CND surrounding 
Sgr~A$^*$, but the local, integrated $P$ is low ($\sim$~1.0\%). Sgr~B2 and Sgr~C show the secondary, brightest polarized flux knots of the map, associated 
with $P$ = 66.5\% for Sgr~B2 and $P$ = 47.8\% for Sgr~C. The difference in $P$ is due to the asymmetrical spatial location of the cloud with respect 
to Sgr~A$^*$. Both the clouds present a polarization position angle $\psi$ perpendicular to the vertical axis of the model (normal to the scattering 
plane). $P$ is found to vary for each CMZ cloud and reaches a maximum for the Eastern and Western sections of the elliptical twisted ring but it is most 
likely that a large fraction of the $\infty$-shaped ring will be diluted by background, unpolarized emission from both plasma emission and Sgr~A$^*$. 
Finally, when integrating the whole 2$^\circ~\times~$2$^\circ$ GC polarized emission, the model produces a net polarization degree of 0.9\% associated with 
$\psi$~=~-22.8$^\circ$. The combined emission from Sgr~A$^*$ and the CND thus dominates the whole polarization picture. 

~\

\citet{Marin2014} extended the modeling of \citet{Churazov2002} by radiatively coupling the primary source to a large panel of reprocessing targets.
They also extend their cloud parametrization to explore the influence of the location of the scattering nebulae on polarization (not shown in this 
research note), finding that the two reflection nebulae always produce high polarization degrees ($\gg$~10\%).

\section{The need for a dedicated X-ray polarimetry mission}
It has been shown by \citet{Churazov2002} and \citet{Marin2014} that the GC is expected to provide a variety of polarization signals where maximum $P$
is expected to arise from reflection nebulae. \citet{Marin2014} also computed the minimum detectable polarization (MDP) that the {\it NHXM} 
(New Hard X-ray Mission, \citealt{Tagliaferri2012a,Tagliaferri2012b}) could have reached and found that such polarization levels are detectable 
using a 500~ks observation. Errors on $\psi$ being marginal, the detection of $\psi$ normal to the scattering plane would be unambiguous.

Was the GC active a few hundreds years ago? X-ray polarimetry can definitively prove or reject this hypothesis since the main molecular clouds should 
be highly polarized ($\gg$~10\%) with the electric vector perpendicular to the line connecting Sgr~A$^*$ to the reprocessing nebula. To spatially
constrain the three-dimensional location of each GC component with respect to the central SMBH, it will be necessary to observe at least two molecular 
clouds. An X-ray mission equipped with a polarization imaging detector, such as the Gas Pixel Detector (GPD) based on the photoelectric effect 
\citep{Costa2001}, is ideally suited since polarization mapping would reveal the complex morphology of the GC, spatially resolving the largest reflection 
nebulae, differentiating them from neighborhood sources and potentially enabling the investigation of stratified light echoes from the past activity of 
Sgr~A$^*$.

A dedicated space mission for imaging X-ray polarimetry such as the non-selected projects {\it NHXM} or {\it IXPE} (the Imaging X-ray Polarimeter Explorer, 
\citealt{Weisskopf2008}) would have been sufficiently sensitive to measure the polarization emerging from the 2$^\circ~\times~$2$^\circ$
GC, even below 8~keV where plasma emission is acting like an unpolarized background and can be further substracted from past spectral data. Finally, 
X-ray polarization measurement would ultimately test the alternative scenario for the origin of X-ray emission from Sgr~B2 and Sgr~C in which X-ray 
features are produced by low-energy cosmic-ray electrons rather than by Compton scattering.

\begin{acknowledgements}
The authors would like to acknowledge the financial support from the COST Action MP1104 and COST-CZ LD12010 .
\end{acknowledgements}

\bibliographystyle{aa}  
\bibliography{marin} 

\begin{thebibliography}{18}
\expandafter\ifx\csname natexlab\endcsname\relax\def\natexlab#1{#1}\fi

\bibitem[{{Bamba} {et~al.}(2002){Bamba}, {Murakami}, {Senda}, {Takagi},
  {Yokogawa}, \& {Koyama}}]{Bamba2002}
{Bamba}, A., {Murakami}, H., {Senda}, A., {et~al.} 2002, ArXiv Astrophysics
  e-prints

\bibitem[{{Churazov} {et~al.}(2002){Churazov}, {Sunyaev}, \&
  {Sazonov}}]{Churazov2002}
{Churazov}, E., {Sunyaev}, R., \& {Sazonov}, S. 2002, \mnras, 330, 817

\bibitem[{{Costa} {et~al.}(2001){Costa}, {Soffitta}, {Bellazzini}, {Brez},
  {Lumb}, \& {Spandre}}]{Costa2001}
{Costa}, E., {Soffitta}, P., {Bellazzini}, R., {et~al.} 2001, \nat, 411, 662

\bibitem[{{Koyama} {et~al.}(1989){Koyama}, {Awaki}, {Kunieda}, {Takano}, \&
  {Tawara}}]{Koyama1989}
{Koyama}, K., {Awaki}, H., {Kunieda}, H., {Takano}, S., \& {Tawara}, Y. 1989,
  \nat, 339, 603

\bibitem[{{Koyama} {et~al.}(2007){Koyama}, {Inui}, {Hyodo}, {Matsumoto},
  {Tsuru}, {Maeda}, {Murakami}, {Yamauchi}, {Kissel}, {Chan}, \&
  {Soong}}]{Koyama2007}
{Koyama}, K., {Inui}, T., {Hyodo}, Y., {et~al.} 2007, \pasj, 59, 221

\bibitem[{{Koyama} {et~al.}(1996){Koyama}, {Maeda}, {Sonobe}, {Takeshima},
  {Tanaka}, \& {Yamauchi}}]{Koyama1996}
{Koyama}, K., {Maeda}, Y., {Sonobe}, T., {et~al.} 1996, \pasj, 48, 249

\bibitem[{{Koyama} {et~al.}(1986){Koyama}, {Makishima}, {Tanaka}, \&
  {Tsunemi}}]{Koyama1986}
{Koyama}, K., {Makishima}, K., {Tanaka}, Y., \& {Tsunemi}, H. 1986, \pasj, 38,
  121

\bibitem[{{Marin} {et~al.}(2014){Marin}, {Karas}, {Kunneriath}, \&
  {Muleri}}]{Marin2014}
{Marin}, F., {Karas}, V., {Kunneriath}, D., \& {Muleri}, F. 2014, \mnras, 441,
  3170

\bibitem[{{Matt}(2010)}]{Matt2010}
{Matt}, G. 2010, {X-ray polarimetry and radio-quiet AGN}, ed. R.~{Bellazzini},
  E.~{Costa}, G.~{Matt}, \& G.~{Tagliaferri}, 122

\bibitem[{{Molinari} {et~al.}(2011){Molinari}, {Bally}, {Noriega-Crespo},
  {Compi{\`e}gne}, {Bernard}, {Paradis}, {Martin}, {Testi}, {Barlow}, {Moore},
  {Plume}, {Swinyard}, {Zavagno}, {Calzoletti}, {Di Giorgio}, {Elia},
  {Faustini}, {Natoli}, {Pestalozzi}, {Pezzuto}, {Piacentini}, {Polenta},
  {Polychroni}, {Schisano}, {Traficante}, {Veneziani}, {Battersby}, {Burton},
  {Carey}, {Fukui}, {Li}, {Lord}, {Morgan}, {Motte}, {Schuller},
  {Stringfellow}, {Tan}, {Thompson}, {Ward-Thompson}, {White}, \&
  {Umana}}]{Molinari2011}
{Molinari}, S., {Bally}, J., {Noriega-Crespo}, A., {et~al.} 2011, \apjl, 735,
  L33

\bibitem[{{Murakami} {et~al.}(2001){Murakami}, {Koyama}, {Tsujimoto}, {Maeda},
  \& {Sakano}}]{Murakami2001}
{Murakami}, H., {Koyama}, K., {Tsujimoto}, M., {Maeda}, Y., \& {Sakano}, M.
  2001, \apj, 550, 297

\bibitem[{{Nobukawa} {et~al.}(2011){Nobukawa}, {Ryu}, {Tsuru}, \&
  {Koyama}}]{Nobukawa2011}
{Nobukawa}, M., {Ryu}, S.~G., {Tsuru}, T.~G., \& {Koyama}, K. 2011, \apjl, 739,
  L52

\bibitem[{{Ponti} {et~al.}(2010){Ponti}, {Terrier}, {Goldwurm}, {Belanger}, \&
  {Trap}}]{Ponti2010}
{Ponti}, G., {Terrier}, R., {Goldwurm}, A., {Belanger}, G., \& {Trap}, G. 2010,
  \apj, 714, 732

\bibitem[{{Sunyaev} {et~al.}(1993){Sunyaev}, {Markevitch}, \&
  {Pavlinsky}}]{Sunyaev1993}
{Sunyaev}, R.~A., {Markevitch}, M., \& {Pavlinsky}, M. 1993, \apj, 407, 606

\bibitem[{{Tagliaferri} {et~al.}(2012){Tagliaferri}, {Hornstrup}, {Huovelin},
  {Reglero}, {Romaine}, {Rozanska}, {Santangelo}, \&
  {Stewart}}]{Tagliaferri2012b}
{Tagliaferri}, G., {Hornstrup}, A., {Huovelin}, J., {et~al.} 2012, Experimental
  Astronomy, 34, 463

\bibitem[{{Tagliaferri} \& {NHXM Consortium}(2012)}]{Tagliaferri2012a}
{Tagliaferri}, G. \& {NHXM Consortium}. 2012, \memsai, 83, 360

\bibitem[{{Weisskopf} {et~al.}(2008){Weisskopf}, {Bellazzini}, {Costa},
  {Ramsey}, {O'Dell}, {Elsner}, {Tennant}, {Pavlov}, {Matt}, {Kaspi}, {Coppi},
  {Wu}, {Siegmund}, \& {Zavlin}}]{Weisskopf2008}
{Weisskopf}, M.~C., {Bellazzini}, R., {Costa}, E., {et~al.} 2008, in {Society
  of Photo-Optical Instrumentation Engineers (SPIE) Conference Series}, Vol.
  7011

\bibitem[{{Yusef-Zadeh} {et~al.}(2002){Yusef-Zadeh}, {Law}, {Wardle}, {Wang},
  {Fruscione}, {Lang}, \& {Cotera}}]{Yusef-Zadeh2002}
{Yusef-Zadeh}, F., {Law}, C., {Wardle}, M., {et~al.} 2002, \apj, 570, 665

\end{thebibliography}

\end{document}